\documentclass[conference]{IEEEtran}
\IEEEoverridecommandlockouts
\usepackage{cite}
\usepackage{amsmath,amssymb,amsfonts}
\usepackage{algorithmic}
\usepackage{graphicx}
\usepackage{textcomp}
\usepackage{xcolor}
\usepackage{enumitem}
\usepackage{dblfloatfix} 
\def\BibTeX{{\rm B\kern-.05em{\sc i\kern-.025em b}\kern-.08em
    T\kern-.1667em\lower.7ex\hbox{E}\kern-.125emX}}

\begin{document}

\title{Dihedral Angle Adherence: Evaluating Protein Structure Predictions in the Absence of Experimental Data\\
}

\author{\IEEEauthorblockN{1\textsuperscript{st} Musa Azeem}
\IEEEauthorblockA{\textit{Department of Computer Science and Engineering} \\
\textit{University of South Carolina}\\
Columbia, United States \\
mmazeem@email.sc.edu}
\and
\IEEEauthorblockN{2\textsuperscript{nd} Homayoun Valafar}
\IEEEauthorblockA{\textit{Department of Computer Science and Engineering} \\
\textit{University of South Carolina}\\
Columbia, United States \\
homayoun@cse.sc.edu}
}

\maketitle

\begin{abstract}
    Determining the 3D structures of proteins is essential in understanding their behavior in the cellular environment. Computational methods of predicting protein structures have advanced, but assessing prediction accuracy remains a challenge. The traditional method, RMSD, relies on experimentally determined structures and lacks insight into improvement areas of predictions. We propose an alternative: analyzing dihedral angles, bypassing the need for the reference structure of an evaluated protein. Our method segments proteins into amino acid subsequences and searches for matches, comparing dihedral angles across numerous proteins to compute a metric using Mahalanobis distance. Evaluated on many predictions, our approach correlates with RMSD and identifies areas for prediction enhancement. This method offers a promising route for accurate protein structure prediction assessment and improvement.
\end{abstract}


\noindent \textit{Regular Research Paper}

\section{Background}
    \subsection{Significance}
        From enabling movement in your body to catalyzing biochemical reactions within your cells, proteins are the building block of life and the backbone behind all biological processes \cite{protein_bg}. Proteins are present in every facet of biology, and understanding the complex functionality of proteins is essential to the advancement of modern medicine. This, however, is not always so simple.
    
        One of the key characteristics of proteins is their three-dimensional shape. Proteins are made up of molecular units known as amino acid residues (residues, for short). All proteins consist of sequences of the 20 amino acid residues present in biology and are identified by their own unique sequence. This sequence determines the protein's three-dimensional structure, which in turn plays a major role in determining the protein's function. In practice, however, the complex interactions between these compounds create intricate structures that are difficult to determine using the protein's amino acid sequence alone. 

    \subsection{Determining Protein Structures}
        Discovering protein structures is an essential step in determining their function in the cellular environment. Naturally, a significant portion of biological research has been dedicated to the discovery of protein structures through experimentation  \cite{protein_market}. In practice, X-ray Crystallography and NMR Spectroscopy are the two primary methods of experimentally determining protein structures \cite{protein_exper}. These methods, albeit slow and expensive, have given way to the $\sim$200,000 protein structures known today \cite{pdb}. 
    
        Given the high cost and time requirements associated with determining protein structures experimentally, modern, computational methods have emerged to bypass this process. In particular, with the advancements of artificial intelligence in recent years, experimentally predetermined structures have been utilized in the training and developing of technologies to predict protein structures using the amino acid sequence alone, circumventing the slow, expensive, and otherwise necessary experimentation \cite{protein_structure_predictions}. AlphaFold, developed by Deepmind at Google, is a landmark instance of these tools, leading the way toward the next stage of protein structure discovery \cite{alphafold, alphafold2}.

\begin{figure*}[b]
    \centering
    \includegraphics[width=0.75\linewidth]{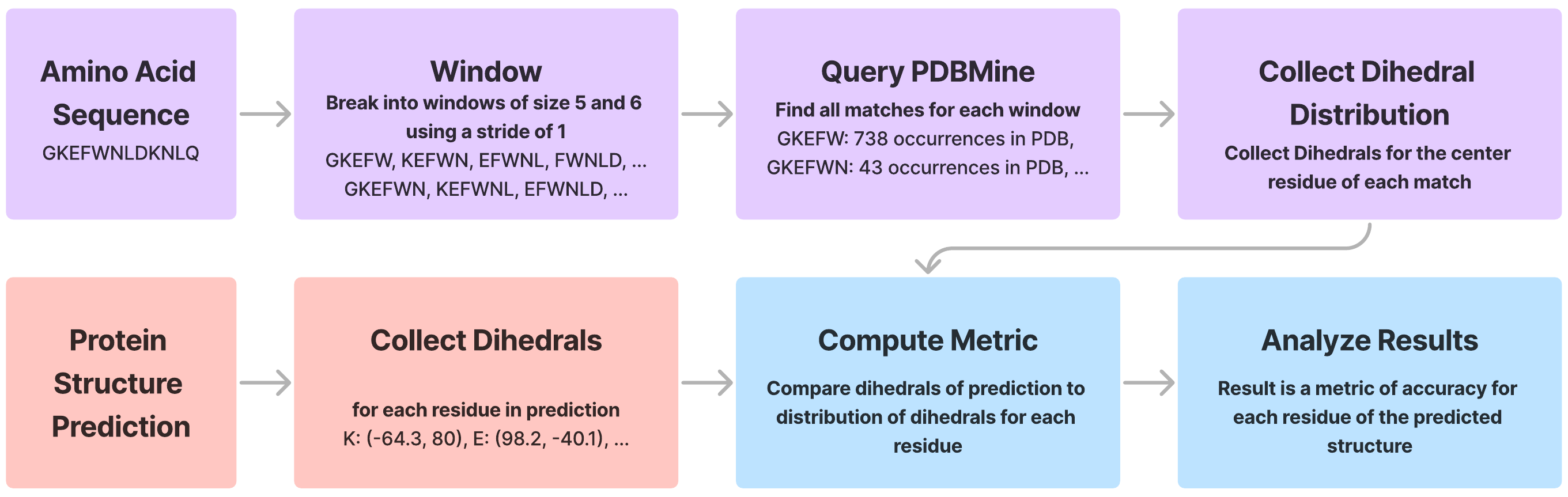}
    \caption{Overview of our methodology. Amino acid sequences and numeric figures are shown as examples}
    \label{fig:method_chart}
\end{figure*}

    \subsection{Representing Protein Structures}
        Discovered protein structures are stored in numerous online databases, with the globally maintained RCSB Protein Data Bank  (PDB) being the most representative \cite{pdb, protein_bg}. Protein structures, whether determined through experimentation or prediction, are typically represented in databases via one of two key attributes. The data most frequently stored in databases such as PDB are the coordinates in 3D space of all heavy atoms of the protein's molecular structure. 

\begin{figure}[t]
    \centering
    \includegraphics[width=1\linewidth]{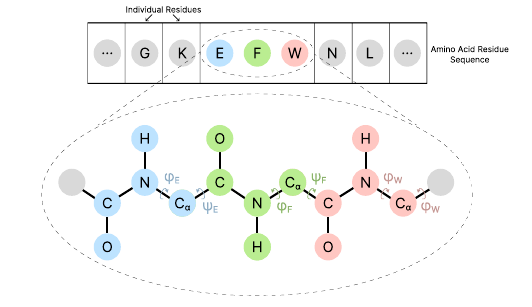}
    \caption{Visualization of the location of dihedral angles, $\phi$ and $\psi$. Example shows the amino acid subsequence of residues E, F, and W. Here, a window size of 3 is illustrated for conciseness.}
    \label{fig:dihedrals}
\end{figure}

        An alternative property capable of representing a protein's structure is its set of dihedral angles \cite{dihedrals}. Illustrated in Figure \ref{fig:dihedrals}, between every residue in a protein's sequence are a pair of bond angles: phi ($\phi$) and psi ($\psi$). Rotations in these angles between residues are the only degrees of freedom allowed for a protein during its formation and are established based on the molecular interactions between residues. As such, these bond angles can be calculated if the 3D coordinates of the protein's molecules are known. Together, the values of these angles at every residue are sufficient to represent the entire protein structure.

\section{Objective}\label{section:introduction}
    A key limitation of the current state of protein structure prediction is the process of evaluating and improving predictions. The current industry standard for evaluating prediction accuracy is Root Mean Square Deviation (RMSD) \cite{rmsd}. This method involves calculating the distance between the corresponding atoms of a prediction and the true structure of a protein, which must be predetermined experimentally. In practice, however, the fundamental goal of predicting protein structures is to circumvent the need for experimentation altogether. To effectively evaluate the performance of protein structure predictions in the absence of the experimentally determined--``ground truth''--structures, an alternative method of evaluation is necessary.
    
    Another limitation of RMSD we address is its lack of insight into which residues in the amino acid sequence of a protein are impacting the overall score most significantly. RMSD produces a single metric for a prediction, representing the adherence of the predicted structure as a whole to the true structure. An alternative metric that provides not just a score, but insight into where a prediction has gone wrong will enable future development in improving protein structure predictions through focus on these locations. 

    In subsequent sections, we discuss our approach and methods for the development of a metric to resolve these issues. Namely, we suggest a method of evaluating protein structures capable of:
    \begin{enumerate}
        \item Effectively evaluating protein structure predictions in the absence of their experimentally determined structures.
        \item Pinpointing where predictions could be improved to facilitate the improvement of these points of prediction.
    \end{enumerate}

    To achieve these goals, we propose examining the dihedral angles of a protein structure rather than its three-dimensional coordinates, as done by methods such as RMSD. We hypothesize that, in general, the dihedral angles of residues appearing in certain amino acid subsequences follow similar patterns across proteins. Consequently, we propose that a predicted structure's dihedral angles adhere to the distribution of angles found across other proteins. Following this approach, we define the following items as a path to achieve the high-level goals for this project:
    
    \begin{enumerate}
        \item Generate a distribution of dihedral angles given amino acid subsequences of a prediction to serve as a baseline for comparison.
        \item Develop a metric to evaluate the performance of a protein structure prediction, comparing its dihedral angles to the baseline distributions.
        \item Extend the tool to determine which points within a prediction perform the worst based on our metric and mark such areas as those that necessitate improvement.
    \end{enumerate}

\begin{figure}[!b]
    \centering
    \includegraphics[width=\linewidth]{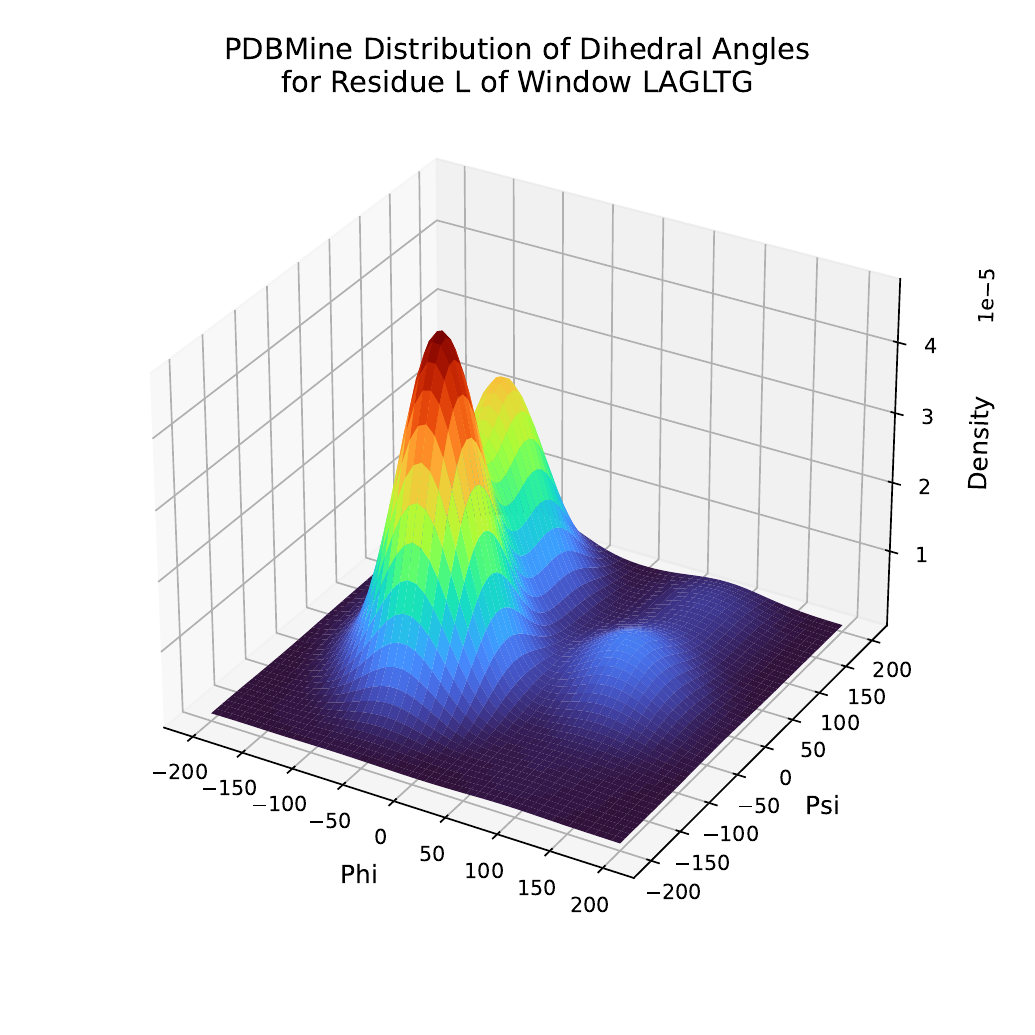}
    \caption{KDE plot of the $(\phi,\psi)$ distribution queried from PDBMine for the subsequence LAGLTG. It is clear that certain values of $\phi$ and $\psi$ are highly probable given this subsequence, while others are near zero.}
    \label{fig:3d_kde}
\end{figure}

\section{Methodology}\label{section:methods}
    Here we discuss our methods of collecting data and computing our proposed metric. An overview of our methodology is shown in Figure \ref{fig:method_chart}.

    \subsection{Data Collection}\label{section:dataCollection}
        As will be explored, a key requirement of our method is the collection of dihedral angles across experimentally determined structures housed in the Protein Data Bank \cite{pdb}. To facilitate efficient querying and collection of dihedral angles, we utilize PDBMine \cite{pdbmine}, an application built on top of PDB to provide an alternative view of its data.

        We initiate the data collection process by windowing the amino acid sequence of a protein with a stride of 1, such that each residue $r_i$ forms a window $W_i$ of size 5 with its neighboring residues. Each window is sent to query PDBMine, which searches the amino acid sequences of all proteins in PDB (other than the protein being evaluated) and returns all occurrences of the window. The dihedral angles of the center residue of every match are extracted, resulting in distributions $f_5^{(i)}(\phi,\psi)$ for every window $W_i$ and corresponding center residue $r_i$ at position $i$ in the amino acid sequence. This collection process is illustrated in Figure \ref{fig:dihedrals}, where the pictured $\phi_F$ and $\psi_F$ are extracted from each match for the example window. In essence, each $f_5^{(i)}$ captures the distribution of dihedral angles for some amino acid (eg. residue $E$ at position $i$) given its window of context (eg. $W_i=\text{GKEFW}$). Figure \ref{fig:3d_kde} demonstrates how certain values of $\phi$ and $\psi$ occur with especially high probability, indicating that predicted values should follow a similar pattern. Finally, this collection process is repeated with a window size of 6, resulting in less populated, but more context-specific, distributions.

\begin{figure}[t]
    \centering
    \includegraphics[width=\linewidth]{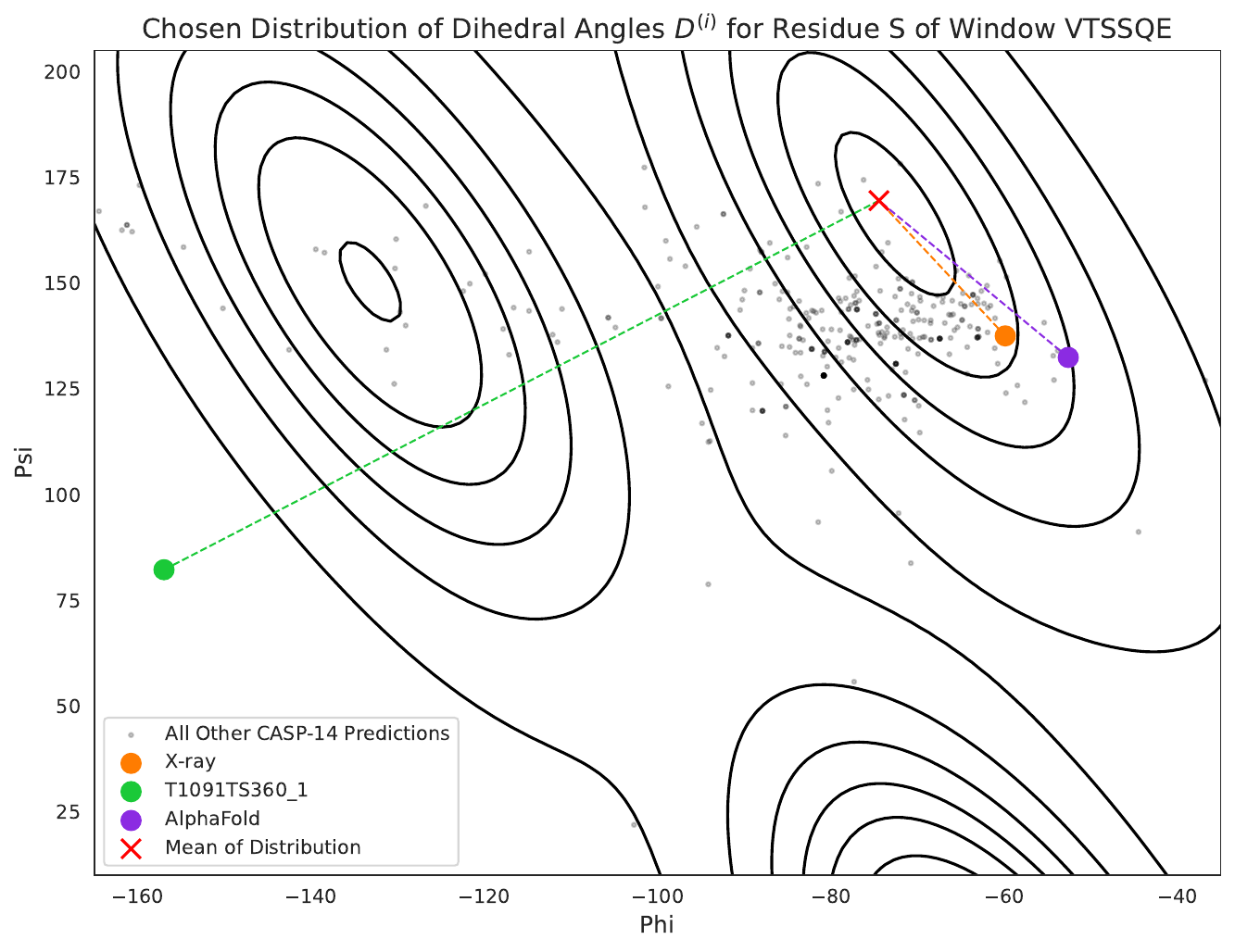}
    \caption{The dihedral distribution chosen as most probable, $D^{(397)}$, for residue S in the protein 7W6B is shown in the KDE plot. Overlaid are $(\phi, \psi)$ data points of interest. The dihedral angles for the X-ray-determined structure at this residue are shown in orange. The same for the AlphaFold prediction, the prediction T1091TS360\_1, and all other predictions submitted to CASP-14 are shown in purple, green, and black, respectively. The mean value of the distribution is shown in red. Dashed lines from each point of interest illustrate the process of calculating the Mahalanobis distance metric. We can see the X-ray-determined structure and AlphaFold's predictions have relatively small distances, while T1091TS360\_1 and some other predictions are very far.}
    \label{fig:md_dist}
\end{figure}

        A similar process is conducted for the protein structure prediction being evaluated. For each amino acid residue $r_i$ in the sequence, the predicted $(\phi_i, \psi_i)$ is extracted, resulting in one dihedral pair per residue, corresponding to the set of distributions found in the previous step. Figure \ref{fig:md_dist} shows one such dihedral pair for a prediction by AlphaFold and the experimentally determined structure overlaid on a KDE plot of the distribution from the previous step. Note that the dihedral angles for the experimentally determined structure serve only as a point of comparison, but are not incorporated into computing our metric.

    \subsection{Computing the Metric}
        The data collection process outlined in Section \ref{section:dataCollection} produces:
        \begin{enumerate}[label=\alph*.]
            \item A predicted angle pair $(\phi_i, \psi_i)$ for each residue $r_i$.
            \item A pair of distributions $f_5^{(i)}(\phi,\psi)$ and $f_6^{(i)}(\phi,\psi)$ of dihedral angles for each residue $r_i$.
        \end{enumerate}
        Now, the following procedure is conducted on each residue $r_i$ in the protein's sequence to produce a per-residue distance metric:
        \begin{enumerate}
            \item \textbf{Kernel Density Estimation} \cite{kde} produces the estimated combined distribution $\hat f^{(i)}(\phi,\psi)$ of $f_5^{(i)}$ and $f_6^{(i)}$ with weights 1 and 128, respectively. 
            \item \textbf{K-Means clustering} \cite{kmeans} on $\{f_5^{(i)}(\phi,\psi),f_6^{(i)}(\phi,\psi)\}$ from PDBMine produces $n$ clusters. $n$ is chosen dynamically to maximize the clustering silhouette score \cite{silhouette}.
            \item The \textbf{most probable} dihedral cluster $D^{(i)}(\phi,\psi)$ is chosen as the cluster that maximizes $\sum\hat f^{(i)}(\phi_j,\psi_j)$ for all points $j$ in the cluster. Following this method, the cluster with the overall most likely dihedral angles is always chosen as the point of comparison.
            \item\textbf{Mahalanobis Distance} \cite{mahalanobis} $M_i$ is computed from the predicted ($\phi_i$, $\psi_i$) to $D^{(i)}$ using the estimated covariance matrix of $D^{(i)}$.
        \end{enumerate}

        The outcome is a distance metric $M_i$ for every residue $r_i$ in the protein's amino acid sequence. We refer to this metric as the \textit{dihedral adherence} of each residue. An illustration of this process for a single window is shown in Figure \ref{fig:md_dist}.

\section{Experiment}
    To test our hypothesis and achievement of goals outlined in Section \ref{section:introduction}, we formulate an experiment involving 4 target proteins of the 14th Critical Assessment of Structure Prediction competition (CASP-14) \cite{casp14}. We look at the following structures:
    \begin{enumerate}
        \item \textbf{Experimentally determined structures}: These protein structures were found through experimentation by X-ray crystallography, and were retrieved from the RCSB Protein DataBank \cite{pdb}. These structures were used in the evaluation of Goal 2.
        \item \textbf{AI-predicted structures}: For each protein, we collected all predicted protein structures submitted to CASP-14, for a total of $\sim$1300 predictions. Of particular interest from this cohort are the predictions of the AlphaFold model \cite{alphafold, alphafold2}.
    \end{enumerate}
    
    For each predicted structure, we follow the procedures outlined in Section \ref{section:methods} to compute the dihedral adherence $M_i$ for each residue.

\section{Analysis \& Results}\label{section:results}
    With a quantified metric of evaluation for each of these predictions, we study two approaches to analyzing our method's achievement of our goals.

\begin{figure}[!b]
    \centering
    \includegraphics[width=\linewidth]{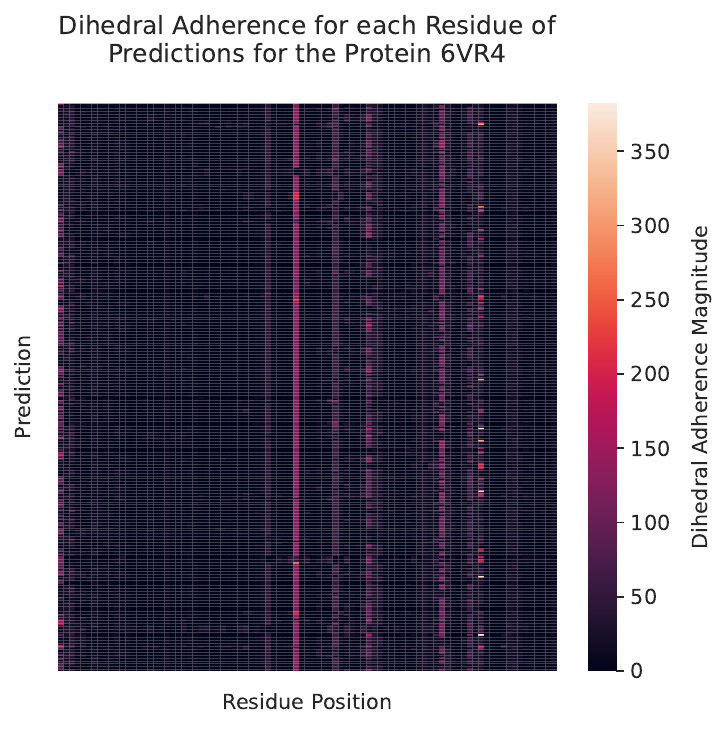}
    \caption{Our calculated dihedral adherence for each residue of each prediction for the protein 6VR4. Variations in certain columns indicate key residues where many predictions disagree.}
    \label{fig:heatmap}
\end{figure}

\begin{figure}[t]
    \centering
    \includegraphics[width=\linewidth]{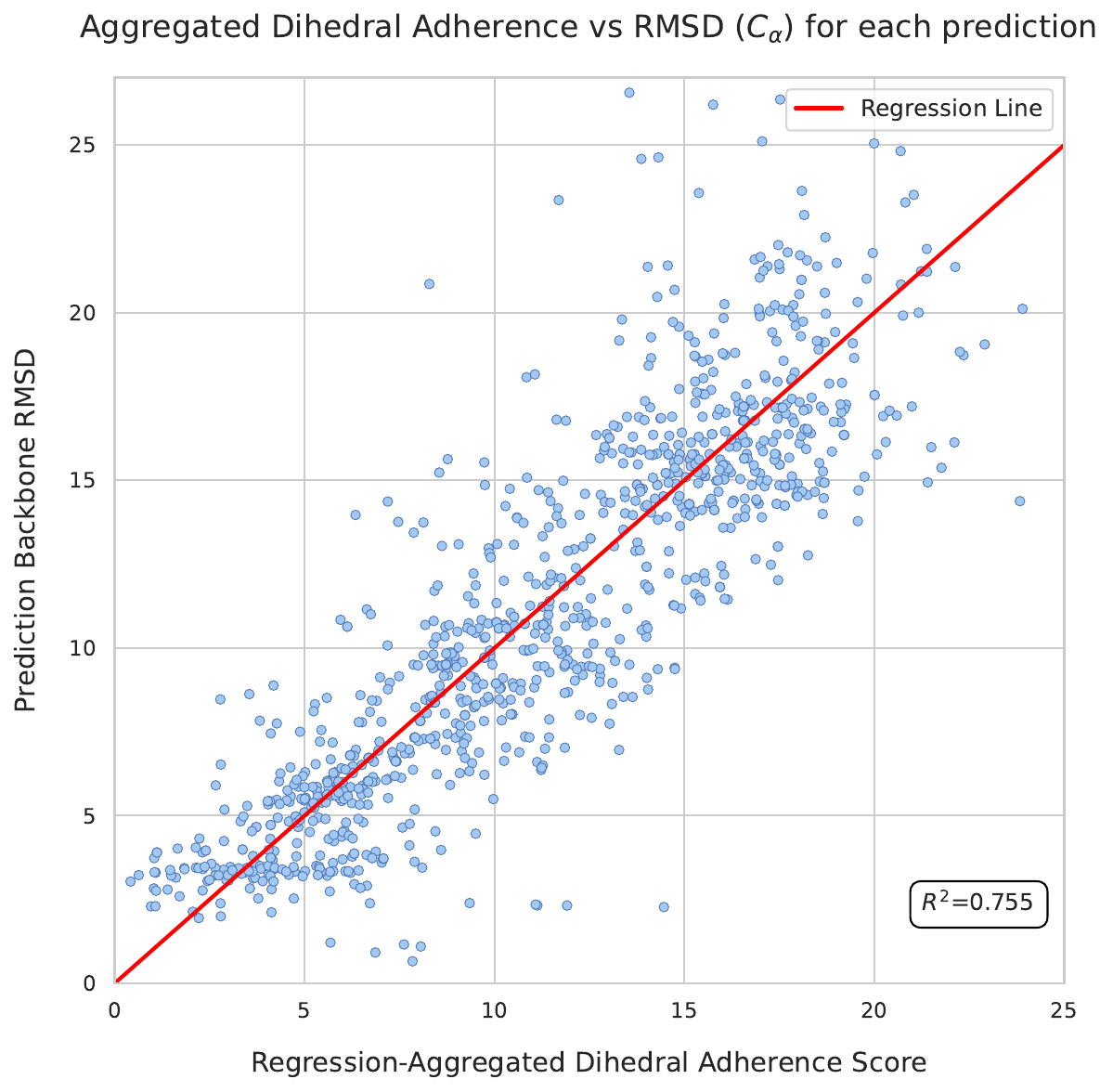}
    \caption{RMSD$_j$ vs $M_j$ for every prediction $j$ submitted to CASP-14 for 4 proteins. Each $M$ is found through the fitted linear regression model. Plotted in red is the line of best fit, with an $R^2$ score of 0.755.}
    \label{fig:MDvsRMSD}
\end{figure}

\begin{figure*}[b]
    \centering
    \includegraphics[width=\linewidth]{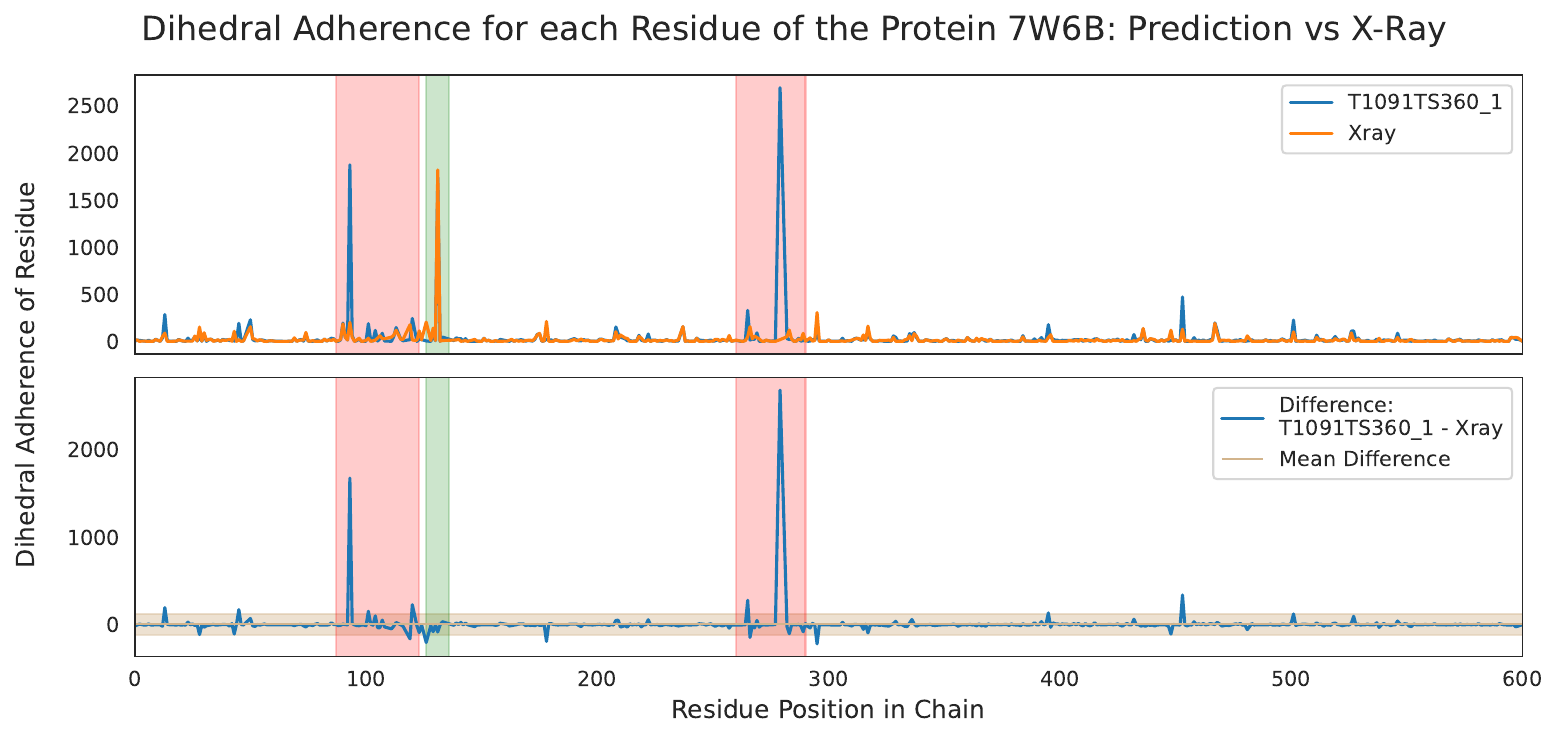}
    \caption{The upper plot pictures the dihedral adherence, $M_i$, for each residue $r_i$ of the protein 7W6B. Our metric is shown in blue for a prediction submitted to CASP-14 (labeled T1091TS360\_1) and in orange for the X-ray-determined structure. The lower plot pictures the per-residue difference ($M_{\text{pred}}-M_{\text{xray}}$) of our metric. The mean difference is shown in tan. On both plots, the regions in red highlight where our metric determines a much greater error in the prediction's performance compared to that of the X-ray structure. On the other hand, the area highlighted in green represents where both the predicted structure and the experimentally determined structure receive a high magnitude of error based on our metric, indicating an area of weakness in our metric rather than a faulty prediction.} 
    \label{fig:ResVsMD}
\end{figure*}

    \subsection{Goal 1: Our Metric as a Viable Method of Evaluating Predictions}
    We first examine the viability of our metric in evaluating protein structure predictions. Illustrated in Figure \ref{fig:heatmap}, we see a variation in dihedral adherence over a multitude of protein structure predictions for the protein 6POO (PDB accession code). We see that there are certain ``key'' residues in this protein for which there is much variation between predictions in the adherence of the dihedrals to the distribution. Based on our hypothesis, predictions with a stronger adherence to the dihedral distribution for these residues are more accurate overall.
    
    To test our hypothesis, we compare our metric to the industry-standard RMSD score of each structure prediction. A correlation between the two indicates that our metric provides insight into the accuracy of a prediction similar to that of the industry standard. To effectively compare the two metrics, we fit a linear regression model \cite{lin_reg} for each prediction from $M$, the set of the 400 dihedral adherences with the greatest magnitudes, to the structure's RMSD. This regression model combines the scores of numerous predictions for 4 proteins submitted to CASP-14, and demonstrates a significant correlation, as presented in Figure \ref{fig:MDvsRMSD}. We see an $R^2$ correlation score of $0.755$ with an F-test p-value $<0.01$.

    We see from these results that there is a significant correlation between our metric and the industry standard of RMSD, demonstrating the viability of our metric in providing similar insight as RMSD without the use of an experimentally determined structure. Although the development of a generalized method of translating our per-residue metric into a single value remains a challenge, the current state of the metric still provides valuable insight into the performance of a protein prediction. In particular, when predicting protein structures that have not yet been discovered experimentally, our metric serves as a viable method of measuring progress to determine which predictions are most optimal. Utilizing this metric in such a way opens the door for future advancements in protein structure predictions.

    \subsection{Goal 2: Pinpointing Areas of a Prediction in Need of Improvement}
    Our method produces a metric of accuracy for every residue within a protein structure prediction. From Figure \ref{fig:heatmap}, we see there exists particular residues in which some predictions perform much better or worse based on our metric. Here, we demonstrate the viability of utilizing these values to pinpoint which residues, in particular, require the most attention to improve a prediction. Ideally, a high magnitude of error computed for a residue would indicate a faulty prediction at this point. For robustness and to account for inaccuracies in our metric itself, however, we compare $M_{\text{prediction}}$ to $M_{\text{true}}$ to see which residues in a prediction produce the highest error \textit{relative} to the error of the X-ray-determined structure. The residues with the greatest discrepancies are highlighted as potential points of interest. 

    Figure \ref{fig:ResVsMD} illustrates our results when comparing a prediction to the X-ray-determined structure for the protein 7W6B. Regions of high discrepancies between the two structures can clearly be identified. These points represent locations where the dihedral angles of a certain residue in a prediction do not closely adhere to the distribution of angles found across other proteins, indicating that something may be off at this location. By comparing the dihedral adherence to that of the experimentally determined structure, we confirm that the predicted angle should ideally be shifted to a more optimal location in the distribution. Following this approach, we identify prime locations in protein structures for investigation in improving predictions in the future.

\section{Future Work \& Conclusions}

    In this paper, we have demonstrated our method's ability to achieve the goals outlined in Section \ref{section:introduction}. The analysis of our per-residue metric of dihedral adherence has high potential in evaluating a protein structure prediction's accuracy. This metric provides similar insight as RMSD, but relies only on predetermined structures in PDB, rather than the experimentally determined structure of the protein being evaluated. Furthermore, by providing a metric for every protein residue, we can identify where in the protein structure a prediction is scoring worse than the experimentally determined structure. These locations serve as prime targets for future improvement of protein structure predictions. Here, we summarize future work that will build upon the foundational methods presented to improve and expand this tool.

    \subsection{Goal 1: Effectively evaluating protein structure predictions:}
        As it is, our metric provides insight into the performance of a protein with a significant correlation to industry-standard methods. However, translating our per-residue metric into a single value comparable to RMSD remains a challenge. In the future, methods of producing a single metric sufficient for evaluating the accuracy of a prediction as a whole will be developed to generalize the linear regression approach outlined in Section \ref{section:results}.

    \subsection{Goal 2: Pinpointing prediction errors:}
        Future work will be directed toward protein structure prediction improvement, utilizing the tools presented here to increase the accuracy of existing predictions. We define the following high-level objectives:
        \begin{enumerate}
            \item Effectively identify residues in a prediction with the worst score without, ideally, comparing them to the ground truth structure.
            \item For such residues, shift the prediction's dihedral angles to a more probable location in the distribution.
            \item After performing energy optimization \cite{energy_opt}, determine if the new structure is more optimal.
        \end{enumerate}

\bibliographystyle{unsrt}
\bibliography{workscited}

\begin{thebibliography}{10}

\bibitem{protein_bg}
David Whitford.
\newblock {\em Proteins: structure and function}.
\newblock John Wiley \& Sons, 2013.

\bibitem{protein_market}
Research and Markets.
\newblock Global 3d protein structure analysis market analysis report 2022: A \$2.67 billion market by 2032 - surging collaborations between major key players to enhance their market presence, Nov 2022.

\bibitem{protein_exper}
Bruce Alberts, Alexander Johnson, Julian Lewis, Martin Raff, Keith Roberts, and Peter Walter.
\newblock {\em Molecular Biology of the Cell. 4th edition.}
\newblock New York: Garland Science, 2002.

\bibitem{pdb}
Christine Zardecki, Shuchismita Dutta, David~S. Goodsell, Robert Lowe, Maria Voigt, and Stephen~K. Burley.
\newblock Pdb-101: Educational resources supporting molecular explorations through biology and medicine.
\newblock {\em Protein Science}, 31(1):129--140, 2022.

\bibitem{protein_structure_predictions}
David Baker and Andrej Sali.
\newblock Protein structure prediction and structural genomics.
\newblock {\em Science}, 294(5540):93--96, 2001.

\bibitem{alphafold}
John Jumper et~al.
\newblock Highly accurate protein structure prediction with alphafold.
\newblock {\em Nature}, 596(7873):583–--589, 7 2021.

\bibitem{alphafold2}
Mihaly Varadi et~al.
\newblock {AlphaFold Protein Structure Database: massively expanding the structural coverage of protein-sequence space with high-accuracy models}.
\newblock {\em Nucleic Acids Research}, 50(D1):D439--D444, 11 2021.

\bibitem{dihedrals}
Saravanan Dayalan, Nalaka Dilshan, Savitri Bevinakoppa, and Heiko Schroeder.
\newblock Dihedral angle and secondary structure database of short amino acid fragments.
\newblock {\em Bioinformation}, 1:78--80, 02 2006.

\bibitem{rmsd}
Oliviero Carugo and Sándor Pongor.
\newblock A normalized root-mean-spuare distance for comparing protein three-dimensional structures.
\newblock {\em Protein Science}, 10(7):1470--1473, 2001.

\bibitem{pdbmine}
Casey Cole, Christopher Ott, Diego Valdes, and Homayoun Valafar.
\newblock Pdbmine: A reformulation of the protein data bank to facilitate structural data mining.
\newblock In {\em 2019 International Conference on Computational Science and Computational Intelligence (CSCI)}, pages 1458--1463, 2019.

\bibitem{kde}
Yen-Chi Chen.
\newblock A tutorial on kernel density estimation and recent advances.
\newblock {\em Biostatistics \& Epidemiology}, 1(1):161--187, 2017.

\bibitem{kmeans}
Anil~K. Jain and Richard~C. Dubes.
\newblock {\em Algorithms for clustering data}.
\newblock Prentice-Hall, Inc., USA, 1988.

\bibitem{silhouette}
Peter~J. Rousseeuw.
\newblock Silhouettes: A graphical aid to the interpretation and validation of cluster analysis.
\newblock {\em Journal of Computational and Applied Mathematics}, 20:53--65, 1987.

\bibitem{mahalanobis}
R.~{De Maesschalck}, D.~Jouan-Rimbaud, and D.L. Massart.
\newblock The mahalanobis distance.
\newblock {\em Chemometrics and Intelligent Laboratory Systems}, 50(1):1--18, 2000.

\bibitem{casp14}
Joana Pereira, Adam~J Simpkin, Marcus~D Hartmann, Daniel~J Rigden, Ronan~M Keegan, and Andrei~N Lupas.
\newblock High‐accuracy protein structure prediction in {CASP14}.
\newblock {\em Proteins}, 89(12):1687--1699, dec 2021.

\bibitem{lin_reg}
Douglas~C Montgomery, Elizabeth~A Peck, and G~Geoffrey Vining.
\newblock {\em Introduction to linear regression analysis}.
\newblock John Wiley \& Sons, 2021.

\bibitem{energy_opt}
Kristin~K. Koretke, Zaida Luthey-Schulten, and Peter~G. Wolynes.
\newblock Self-consistently optimized energy functions for protein structure prediction by molecular dynamics.
\newblock {\em Proceedings of the National Academy of Sciences}, 95(6):2932--2937, 1998.

\end{thebibliography}

\end{document}